\documentclass[multphys,vecphys]{svmult}

\usepackage{makeidx}         
\usepackage{graphicx}        
\usepackage{multicol}        
\usepackage[bottom]{footmisc}

\begin{document}

\title*{The low-luminosity galaxy population in the NGC\,5044 Group}
\titlerunning{Low-luminosity galaxies in the NGC\,5044 Group}

\author{Sergio A. Cellone\inst{1}\and
Alberto Buzzoni\inst{2}}

\authorrunning{S. A. Cellone \& A. Buzzoni}

\institute{Fac.\ de Cs.\ Astron\'omicas y Geof\'{\i}sicas (UNLP), Paseo del
  Bosque, B1900FWA La Plata, Argentina
\texttt{scellone@fcaglp.unlp.edu.ar}
\and INAF -- Osservatorio Astronomico di Bologna, Via Ranzani 1, Bologna,
  Italy \texttt{alberto.buzzoni@bo.astro.it}}

\maketitle
\begin{abstract}
 Detailed surface photometry for 79 (mostly dwarf) galaxies in the NGC\,5044
 Group area is analysed, revealing the existence of different morphologies
 among objects originally classified as early-type dwarfs. Particularly, a
 significant fraction of bright dwarf ``ellipticals'' show a distinct
 bulge+disc structure; we thus re-classify these objects as dwarf
 lenticulars (dS0). Our finding points at a possible scenario where these
 systems are the remnants of ``harassed'' disc galaxies. This is emphasized
 by the discovery of a few objects with hints for very low-surface
 brightness spiral-like structure.  The colours, structure, and spatial
 distribution of the different galaxy types suggest that our classification
 may indeed be separating objects with different origins and/or evolutionary
 paths.
 \end{abstract}

\section{Introduction and observational material}

Due to their low luminosities and sizes, a detailed classification of dwarf
($M_B \ge -18$) galaxies is not easy, thus leading to a deceptively simple
picture: those objects showing conspicuous signatures of present/recent star
formation and interstellar material are called dwarf irregulars (dI), while
the remaining smooth-looking, gas-poor objects fall within the dwarf
elliptical (dE) designation%
\footnote{Blue compact dwarfs (BCD), at least in groups and clusters,
are rare objects, and will not be treated here.}%
. However, there is growing evidence for a morphological diversity among
dwarfs; in particular, embedded disc structure and/or rotation were
discovered within a fraction of dEs \cite{cell:JKB00, cell:PGCSG02,
cell:SP02, cell:dRDZH03, cell:GJG03}, which seems to favor the idea that a
fraction of the dEs may be remnants of ``harassed'' disc galaxies
\cite{cell:MLK98}. Whether these objects are related to the still poorly
known class of dwarf lenticulars (dS0) or not, is still matter of debate
\cite{cell:AIVMS05, cell:LGB05}.
 
Aiming at a comprehensive study of the low-luminosity galaxy population in
the NGC\,5044 Group ($m-M=31.9)$ \cite{cell:FS90} we have gathered multicolour
surface photometry for a representative sample, comprising 40 galaxies with
Gunn system $griz$ imaging data, observed with the ESO 3.6-m telescope
(1999--2000), and 57 galaxies observed at CASLEO, Argentina (1996--1999) in
$V$ and $B$ or $R_\mathrm{C}$ with a 2.1-m telescope. There are 18 objects
in common between both subsamples, hence we have a total of 79 different
galaxies on the NGC\,5044 Group area, observed at least in one photometric
band. Of these, 74 galaxies have at least one colour information. A
subsample of 13 galaxies was also observed spectroscopically at ESO. First
results, involving nearly 50\% of these data, have been already presented
\cite{cell:C99, cell:CB01, cell:CB05}.

\begin{figure}[!t]
\setlength{\fboxsep}{0pt}
\centering
\hfill \includegraphics[height=4.5cm]{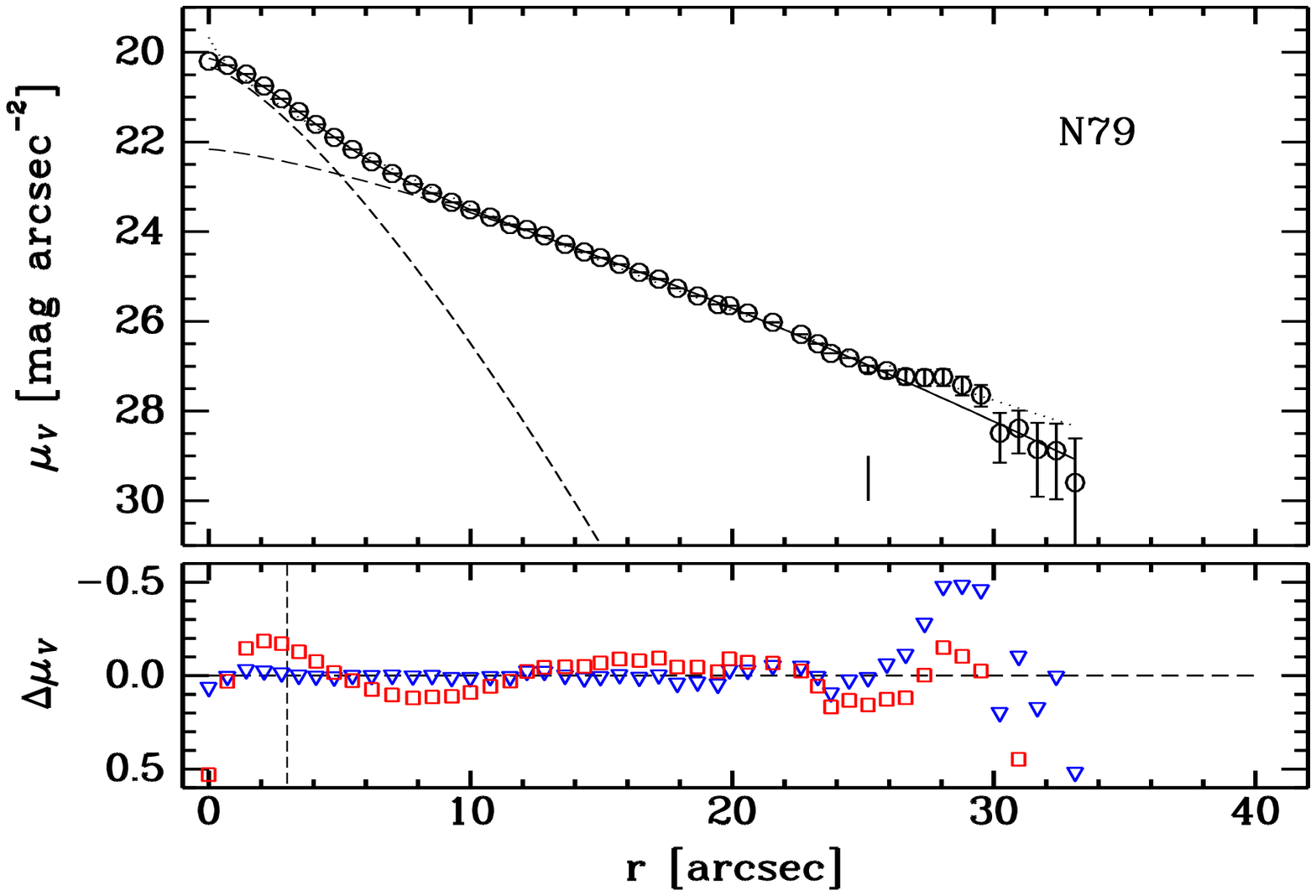}%
 \hfill\fbox{\includegraphics[height=4.5cm]{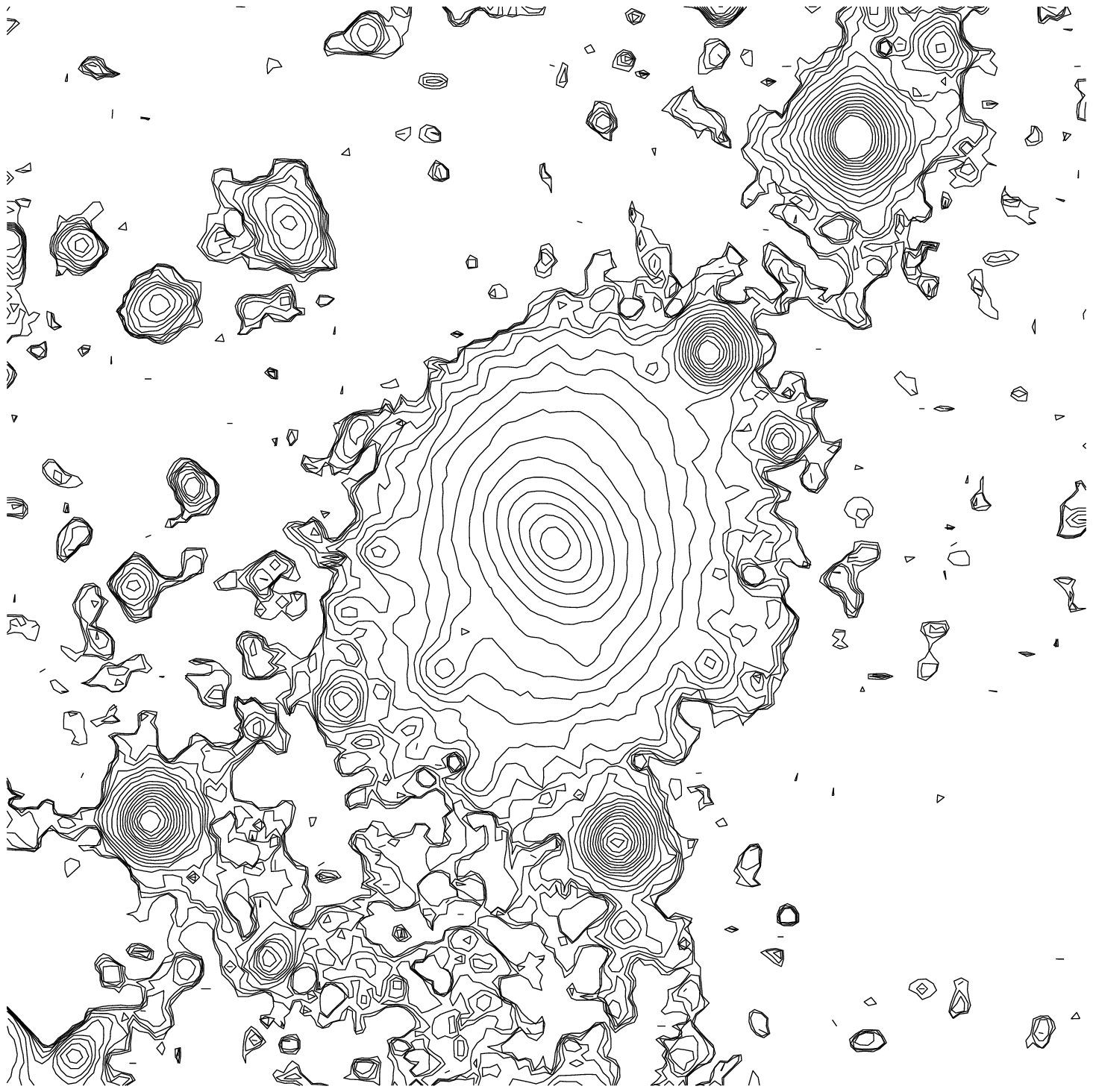}}\hspace*{\fill}
\caption{\emph{Left -} Surface brightness profile for N79 (dS0) showing
  S\'ersic fits to the whole profile and to bulge and disc components
  (dashed lines); the solid line is the B+D fit. The lower panel shows
  residuals from a S\'ersic fit to the whole profile (squares) and from
  the B+D fit (triangles). \emph{Right -} Contour plot for N79.}
\label{cell:figdS0}
\end{figure}

\section{Classification}

Background objects were identified by means of morphological criteria when a
redshift was not available (see \cite{cell:CB05}).  Group members were
classified mostly relying on the behaviour of their surface brightness
profiles (SBP), with the aid of colour information and morphological
appearance. We were able to assign any individual Group member into one of
the following classes:
\begin{description}
\item[dE:] A S\'ersic law \cite{cell:S68} gives very good fits to their
SBPs. Generally, these galaxies show no isophote twisting.
\item[dE/dS0:] These objects are well fit by bulge+disc (B+D) models. They
usually show isophote twisting, ellipticity gradients, and/or colour
gradients.
\item[dI/dE:] Very low surface brightness (LSB) objects, with very extended
and nearly exponential SBPs (S\'ersic index $n \simeq 1$).
\item[Im:] ``Magellanic'' irregulars.
\item[dSph:] Objects (mostly new ones) with $M_g \ge -12$ and central
surface brightnesses $\mu_0 \ge 24$ mag arcsec$^{-2}$.
\end{description}

As an example, Fig.~\ref{cell:figdS0} shows the SBP and contour plot for an
object (N79) we re-classify as dS0. Trying a single S\'ersic fit to the
whole useful profile leaves both positive and negative systematic residuals
(``wave pattern,'' see \cite{cell:BGDP03}). The contour plot (right) clearly
shows this galaxy's isophote twisting.

The dI/dE class, in turn, includes a few objects with LSB outer spiral arms
\cite{cell:CB05}.

\begin{figure}[!t]
\centering
\includegraphics[width=0.495\hsize]{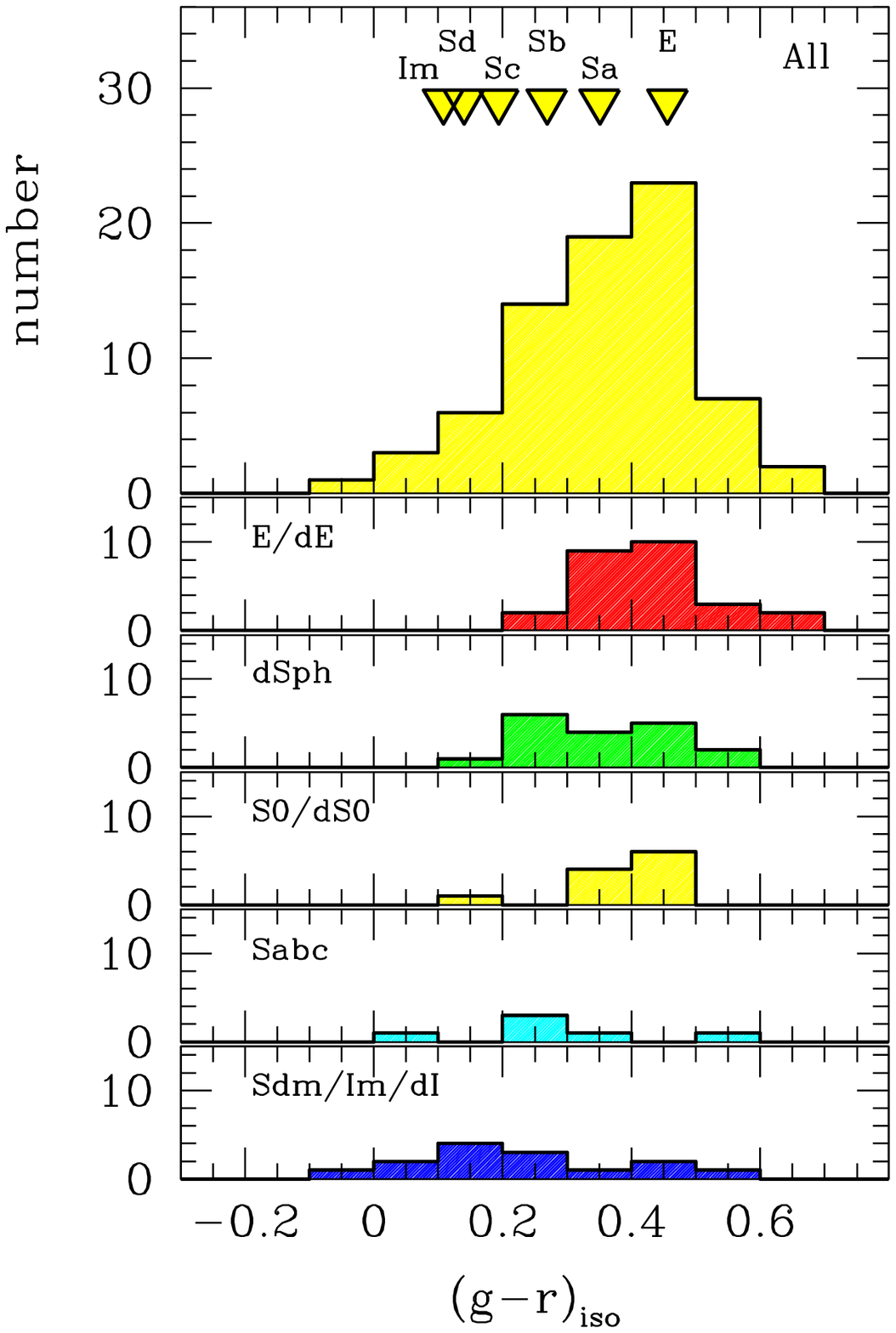}%
\hfill\includegraphics[width=0.495\hsize]{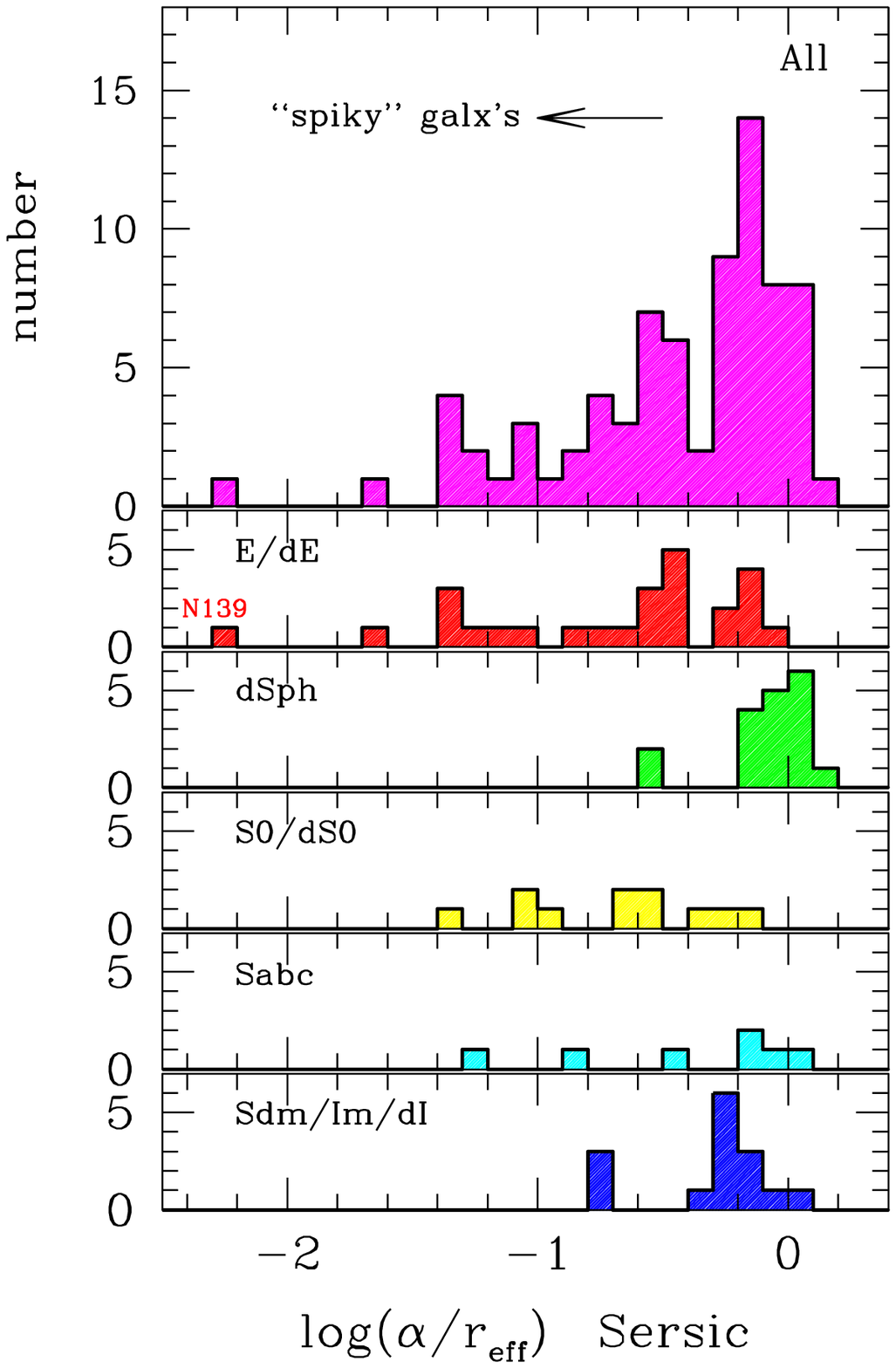}
\caption{\emph{Left -} Galaxy colour distribution for the whole sample (top)
and for each morphological type.  The integrated ($g-r$) colour refers to
$g$ and $r$ galaxy luminosity collected within a $\mu(g) =
27$~mag\,arcsec$^{-2}$ isophotal aperture.  The 15 Gyr template galaxy
models from \cite{cell:B02,cell:B05} are compared for reference (top
triangle markers, as labeled in the upper panel).  \emph{Right -} Same as
left for the ``compactness'' parameter $\log(\alpha/r_\mathrm{eff})$,
defined as the ratio between S\'ersic pseudo scale-length and galaxy
effective radius; compactness increases from right to left. Note the
outstanding case of galaxy N139, likely a background cD at $z \sim 0.4$.}
\label{cell:fcolhist}
\end{figure}

\begin{figure}[!t]
\begin{minipage}{0.58\hsize}
\includegraphics[width=0.94\hsize]{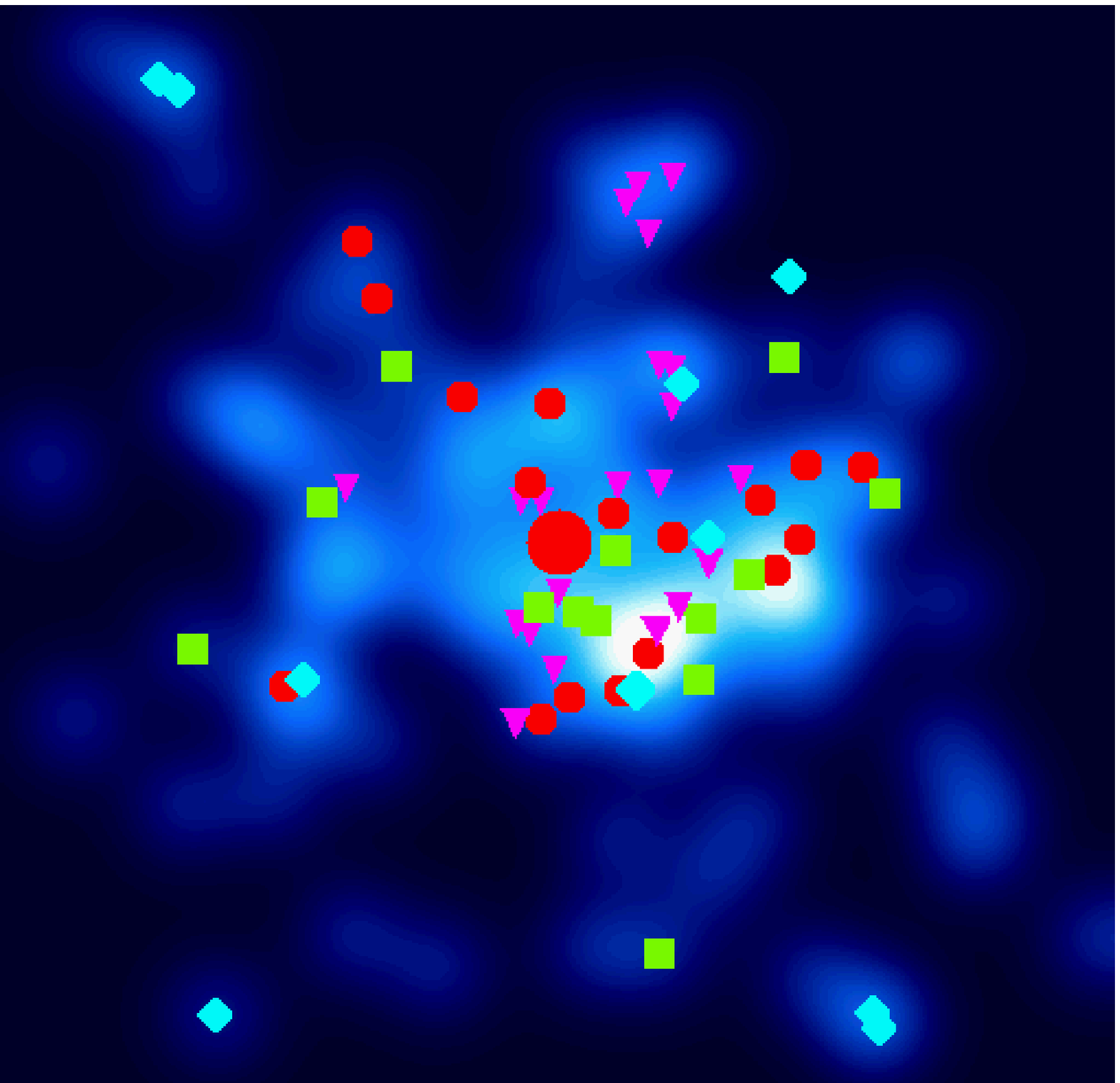}
\end{minipage}
\hfill
\begin{minipage}{0.42\hsize}
\caption{\emph{Left panel -} Projected number-density map, considering only
definite and likely member galaxies from \cite{cell:FS90}.  Superposed is the
galaxy distribution from our sample: E+dE (circles); S0+dE/dS0 (squares);
S--Im+dI/dE (diamonds); dSph (triangles). The big central dot is
NGC\,5044. 
\emph{Lower panels -} Galaxy distribution for different morphological types 
against distance to the nearest bright ($B_\mathrm{T} < 15$ mag) galaxy.
Note the striking spatial segregation of dwarf spheroidals, preferentially
located around bright group members.}
\end{minipage}
\includegraphics[width=\hsize]{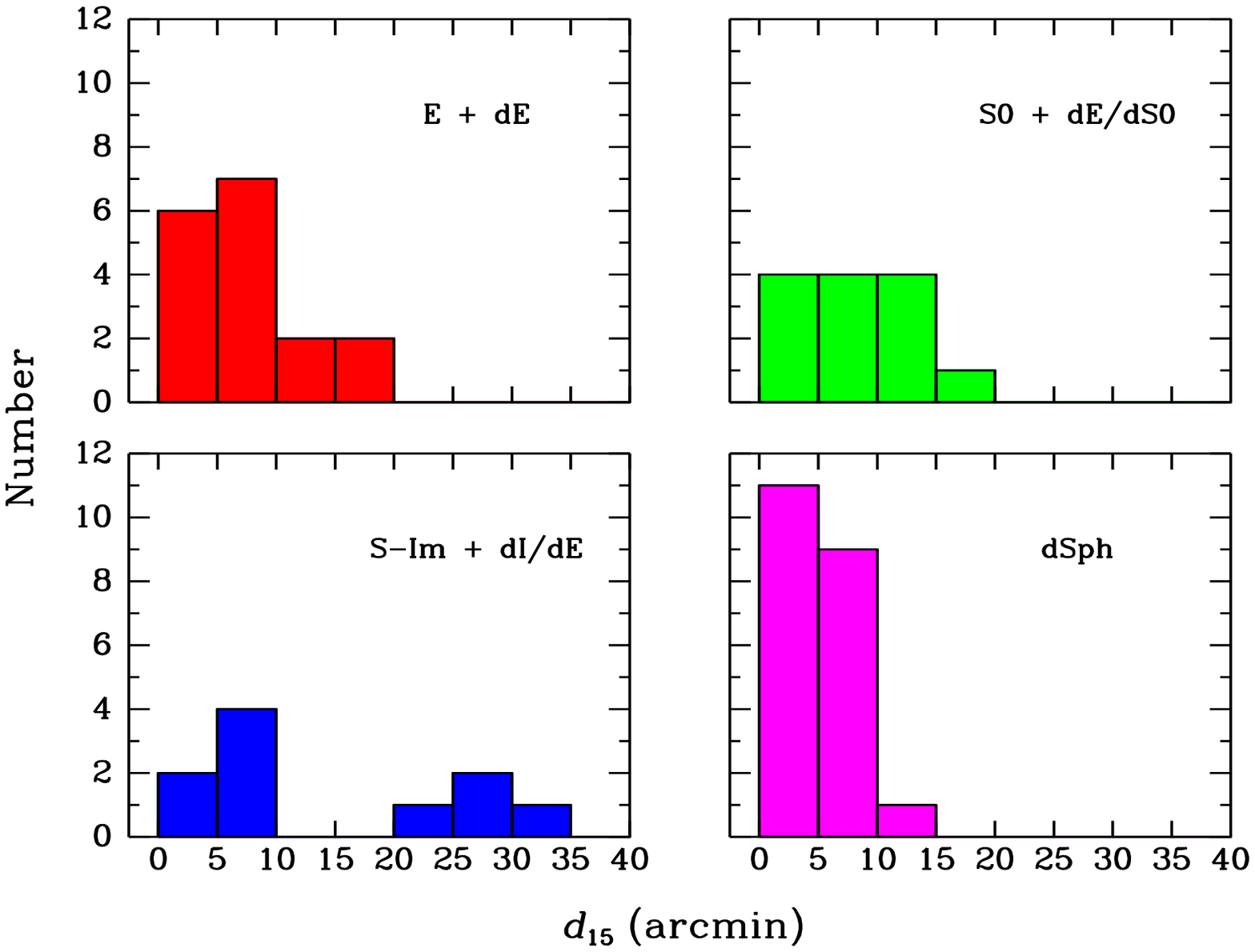}
\label{cell:figdens}
\end{figure}
\section{Photometric properties}
The galaxy colours in our sample show the usual trend with
morphological type, with later types having bluer mean colours
(Fig.~\ref{cell:fcolhist}, left). The blueing from dE's to dSph's, in turn,
is most probably due to a luminosity--metallicity relation.  Dwarf
spheroidals display a flat and mildly broad distribution, due in part to
photometric errors (worse for these faint objects), but probably
also reflecting an intrinsic scatter in their origins and star formation
histories, as is known for their Local Group counterparts (see
e.g.~\cite{cell:GGH03}). Also the latest types have a broad color distribution,
although in this case the cause may be internal reddening.

Structural differences between morphological types may be tested by means
of a compactness parameter defined as $\alpha/r_\mathrm{eff}$, where $\alpha$
is the pseudo scale-length in S\'ersic's formula, and $r_\mathrm{eff}$ is the
effective radius.  Fig.~\ref{cell:fcolhist} (right) shows the distributions
of $\log(\alpha/r_\mathrm{eff})$ for each morphological type. The 
dSph galaxy class is characterized by a shallow SBP, clearly distinct
from the dE one (but see \cite{cell:CB05} for possible selection
effects). Among the latter, N139 (the most compact object labelled on 
Fig.~\ref{cell:fcolhist}) stands out for its extremely ``spiky''
SBP; our photometric redshift estimate locates this (likely cD) galaxy in 
the far background at $z\simeq 0.4$.

\section{Projected spatial distribution}
There is evidence for a morphology -- density relation within the NGC\,5044
Group, as shown in Fig.~\ref{cell:figdens}. While the S0+dE/dS0 galaxies are
intermediate between E+dE (more concentrated) and late-type objects (less
concentrated), dSph's seem to prefer the highest density regions in the
Group. Differences between galaxy types become marginally more
significant when the distance to the nearest bright ($B_\mathrm{T} < 15$
mag) neighbour is considered instead of local number density (see
lower panels in Fig.~\ref{cell:figdens}). In fact, dSph's are only found 
in the (projected) vicinity of brighter member galaxies.

\begin{acknowledgement}
SC would like to thank the conference organizing committee.
This project received partial financial support from CONICET (Argentina) and
the Italian INAF, under grant PRIN/05.
\end{acknowledgement}



\end{document}